 \documentclass[12pt,preprint]{aastex}

 \usepackage{emulateapj5}

\slugcomment{to appear in the Astrophysical Journal, Letters}

\shorttitle{V445 Pup, a He nova}
\shortauthors{Kato and Hachisu}

\begin{document}


\title{V445 Puppis  -- Helium Nova on a Massive White Dwarf}


\author{Mariko Kato}
\affil{Department of Astronomy, Keio University, Hiyoshi, Yokohama
223-8521, Japan}
\email{mariko@educ.cc.keio.ac.jp}

\and

\author{Izumi Hachisu}
\affil{Department of Earth Science and Astronomy, College of Arts and 
Sciences, University of Tokyo, Komaba, Meguro-ku, Tokyo 153-8902, Japan}
\email{hachisu@chianti.c.u-tokyo.ac.jp}


\begin{abstract}
The 2000 outburst of V445 Puppis shows unique properties, such as absence
of hydrogen, enrichment of helium and carbon, slow development of 
the light curve with a 
small amplitude that does not resemble any classical novae.
This object has been suggested to be the first example of helium novae. 
We calculate theoretical light curves of helium novae and reproduce
the observational light curve of V445 Pup.
Modeling indicates a very massive white dwarf (WD), more massive than 
$1.3~M_\sun$. The companion star is possibly either a 
helium star or a helium-rich main-sequence star. We estimate 
the ignition mass as several times $10^{-5} M_\sun$, the corresponding 
helium accretion rate as several times $10^{-7} M_\sun$~yr$^{-1}$,
and the recurrence period as several tens of years. 
These values suggest that the WD is growing in mass and ends up 
either a Type Ia supernova or an accretion induced collapse 
to a neutron star.
\end{abstract}



\section{Introduction}
The outburst of V445 Puppis was discovered on 30 December 2000
by Kanatsu \citep{kan00}. The spectrum shows absence of hydrogen, 
enrichments of helium and carbon. Optical spectrum is not yet published 
in full papers, but near infrared spectra confirm the absence of 
Paschen and Brackett hydrogen lines \citep[][and references therein]{ash03}.
Light curves reported by VSNET  
\footnote{VSNET:http://vsnet.kusastro.kyoto-u.ac.jp/vsnet/index-j.html}
show its small amplitude of the outburst ($\Delta m_{\rm v} \sim 6$) 
and the shape of light curve which do not resemble those of classical 
novae, recurrent novae or other cataclysmic variables.
From these unique properties, this object is suggested to be a helium 
nova \citep{ash03}. 

Infrared observations suggest dust formation. 
During the outburst, infrared spectra consistent with dust emission 
are observed on 2001 Jan 2, JD~2,451,912  \citep{ash03}, and 
on 2001 Jan 31, JD~2,451,941 \citep{lyn01}.
\citet{ash03} argued that the thermal emission comes from dust
formed during the 2000 outburst, while \citet{lyn01} claimed that
it comes from a preexisting dust shell of a previous outburst.  
\citet{hen01} reported that V445 Pup was fainter than $V \sim 20$  
on 2001 October 4, JD~2,452,187, with a remark that the object is 
evidently shrouded by a thick and dense carbon dust shell. 
\citet{ash03} reached the same conclusion based on the $JHK$ observation 
on 2001 November 1, JD~2,452,215. To summarize, an optically thin dust 
shell exists on the outburst phase, and a thick dust shell is formed 
80 days  after the star fades away.

Helium novae have been theoretically predicted by \citet*{kat89} as a
nova outburst phenomenon caused by helium shell flash 
on a white dwarf (WD). They considered two cases of helium accretion;
a helium accretor from the companion helium 
star, or a hydrogen-rich matter accretion  from a 
normal companion with a high accretion
rate. In the latter case, a part of hydrogen-rich matter accreted is 
processed into helium and it accumulates on the WD. 
When the mass of the helium layer reaches a critical mass, an unstable
weak helium shell flash occurs. This is the helium nova. 
In a helium nova, mass loss owing to optically thick wind is relatively weak,
and most of the helium envelope burns into carbon and oxygen,  
and accumulates on the
WD \citep{kat99}. Through many periodic helium shell flashes, 
the WD gradually grows in mass and ends up an accretion induced collapse to
a neutron star or explodes as a Type Ia supernova \citep*{nom84}.

In this $letter$, we present light curve modeling  of 
V445 Pup and show that this object is indeed a helium nova on 
a massive WD. We also estimate the helium accretion rate 
and the recurrence period of outbursts.


\section{Input Physics}

The decay phase of helium nova outbursts can be followed 
by using an optically thick wind theory \citep{kat94}. 
The structure of the
WD envelope is calculated by solving the equations of motion, 
continuity, energy transport, and energy conservation. 
The details of computation have been already presented 
in \citet{kat99} for helium shell flashes on a $1.3~M_\sun$ WD.
Here, other various WD masses are examined that are 
1.1, 1.2, 1.3, 1.33, 1.35, and 
$1.377~M_\sun$.  For the 1.1 and 1.2 $M_{\odot}$ WDs, we assume the 
Chandrasekhar radius,
and for the 1.3 $M_{\odot}$ WD, we adopt the radius of helium burning zone,
 $\log R_{\rm WD}=8.513$, in \citet{kat89}. For more massive WDs, we adopt
the radius of accreting WDs \citep{nom84}. 
The $1.377~M_\sun$ WD is a limiting case of mass-accreting WDs
just before a central carbon ignition \citep{nom84}.
The composition of the envelope is assumed to be uniform
with $(X, Y, C+O , Z) = (0.0, 0.48 ,0.5, 0.02)$ for $\le 1.3~M_\sun$,
which  is taken from \citet{kat99}, 
and $(0.0, 0.38, 0.6,0.02)$ for $\ge 1.33  M_\sun$.
Changing the ratio of carbon to oxygen with the total mass ratio
constant $(C+O = $const.) hardly changes the result.  OPAL opacity is used.

\section{Light Curve Modeling}

Figure 1 shows the photospheric temperature $T_{\rm ph}$, 
the photospheric wind velocity $V_{\rm ph}$, the photospheric 
radius $R_{\rm ph}$,  
the wind mass loss rate and the total mass decreasing rate of the envelope
(the wind mass loss rate plus the nuclear burning rate). At the 
maximum expansion, 
the star reaches somewhere on the curve depending on the envelope mass, 
and moves leftward in time. The wind mass 
loss stops at the point marked by the small open circles. After that  
the star further moves 
leftward owing to nuclear burning.  
The solution of $1.3~M_\sun$ is already
presented in \citet{kat99}.

Figure 2 shows theoretical light curves corresponding to the models in 
Figure 1. A more massive WD shows a more rapid decline because of 
a smaller helium envelope mass as shown in Figure 1. 
Difference in the chemical composition, for example,
$(Y, C+O)=(0.38,0.6)$ and $(0.48, 0.5)$ on the $1.3~M_\sun$ WD,
hardly changes the theoretical light curves as shown in Figure 2
(see the difference between the dotted curve and the solid 
curve). 

After the onset of a helium shell flash the star brightens up and 
reaches somewhere on the theoretical curve and moves rightward on the 
curve in time (Figure 2). Note that the $t=0$ in these template light curves is 
not the onset of ignition of a particular object. The first point is 
determined by the ignition mass, i.e., the envelope mass at helium 
ignition for an individual object. If it is more massive, it first 
appears much lefter-side on the light curve, and the outburst lasts 
for a longer time. The $V$-magnitude decreases slowly in the early stage 
and goes down quickly in the later stage. Especially after the wind stops at 
the point denoted by asterisk, the star quickly fades away. The helium 
burning stops at the last point of each curve.

Observational data of V445 Pup, taken from VSNET, is shown in Figure 3. 
The mean decreasing rate of V445 Pup is 
about $1.8/200$ (mag~day$^{-1}$). 
The light curves of the 1.1 and $1.2~M_\sun$ WD decline too slow
 to be compatible with the observation  
even if we choose any parts of their light curves. 
On the other hand, 
we are able to fit a part of our theoretical light curves
for more massive than $1.3~M_\sun$ WD to the observational data.
Thus, we may conclude that V445 Pup has a
WD more massive than $1.3~M_\sun$. 

To fit the vertical axis of the light curves, we obtain 
an apparent distance modulus of $(m-M)_V = 9.8$
as shown in Figure 3.  The horizontal axis (time) is also
shifted to fit the data.  Here we adopt
a middle part of our theoretical light curves around $M_V 
\sim 1$ that has a decline rate of $\sim 1.8/200$ (mag~day$^{-1}$).

A slow decline lasts longer than 200 days 
and the brightness suddenly drops. 
This point corresponds to the cease point of the optically thick wind.
After the wind mass loss stops, the photosphere shrinks very rapidly. 
This feature is also shown in theoretical light curves for 
classical novae \citep{kat97} and 
recurrent novae \citep{kat99rn} of very massive WDs. 
This also appears in theoretical light curves of helium novae.

All the theoretical curves of $1.2 - 1.377~M_\sun$ are almost similar 
until JD~2,452,100, and deviate from each other after that day.
The light curve of 1.1 $M_{\odot}$ is too slow to decline
and it should be excluded in the following discussion. 
We also think that the 1.2 and $1.3~M_\sun$ WDs do not meet 
the upper limit (arrows) of the observational data after JD~2,452,100.
Finally, we adopt the conclusion that V445 Pup has a massive 
($\ge 1.33~M_\sun$) WD.

\section{Discussion and Conclusions}

V445 Pup shows a quick decrease in the light curve 
from JD~2,452,100 in Figure 3. One may attribute this quick decrease  
to a thick dust shell formation in carbon-rich ejecta. 
If a dust shell is quickly formed and becomes optically thick to block 
the stellar light, we may expect a quick decrease in the light curve. 
In this case, the star moves along the theoretical curve 
in Figure 3 for a while, and then drops down 
earlier than the theoretical fade-away.
Even if a black-out by dust occurs in this system,
the light curve modeling during the slow decline phase does hardly change. 
For a WD less massive than $1.2~M_\sun$, 
any part of light curves cannot fit the data and is also excluded.
For the $1.3~M_\sun$ WD (and barely the $1.2~M_\sun$ WD), 
it is  required that a thick dust shell formation effectively occurs 
just near the theoretical fade-away to meet the observational 
data (the upper limits, arrows in Fig. 3).
For more massive WDs than $1.3~M_\sun$, 
we can shift the theoretical light curve rightward, but it is 
as large as by about 100 days from the limit of the decline rate,
$\sim 1.8/200$ (mag~days$^{-1}$).
Therefore, we safely conclude that the WD of V445 Pup 
is still as massive as $1.3~M_\sun$ or more.

The recurrence period of helium novae is estimated as follows:
The ignition mass, i.e., the envelope mass at the onset of helium shell 
flash is approximated by the mass of our wind solution at the optical
peak. As the beginning time of eruption is not certain, we 
assume that it begins at JD~2,451,820 in Figure 3, i.e., shortly after 
the latest upper limit observation at JD~2,451,814. 
The schematic light curve 
is shown by the vertical line in Figure 3. Then the 
ignition mass is approximated by the mass of the wind
solution marked by the square, 
which is $4.9 \times 10^{-5} M_\sun$ for $1.33~M_\sun$ WD.
A model of helium-shell flashes predicts a relation between the
mass accretion rate and the ignition mass. According to Saio's
numerical calculation of steady helium accretion (Saio 2003, private 
communication), the corresponding mass accretion rate is
$7.1 \times 10^{-7} M_\sun$~yr$^{-1}$. 
The recurrence period of 
helium novae is simply estimated, from the ignition mass divided 
by the accretion rate, as 69 years. If we take the beginning of 
the outburst as the date of discovery, JD~2,451,868, 
these values slightly changed to be $4.6 \times 10^{-5} M_\sun$, 
$7.6 \times 10^{-7} M_\sun$~yr$^{-1}$, and 61 years, respectively. 
For the case of a $1.35~M_\sun$ WD at JD~2,451,820,  
they are  $4.2 \times 10^{-5} M_\sun$,
$5.0 \times 10^{-7} M_\sun$~yr$^{-1}$ and 84 years.

The distance to the star has been poorly known.  We have estimated from 
the comparison of the absolute magnitude of theoretical light curves 
and the apparent magnitude of observed data to be 
about 640 pc for $1.33 M_{\odot}$ and 700 pc for  $1.35 M_{\odot}$ 
with an assumed extinction A$_{\rm v}=0.78$ \citep{ash03}. 
This value is consistent with an upper limit of 3 kpc based on the
strength of an interstellar absorption given by Wagner 
(http://vela.as.arizona.edu/~rmw/v445pup.html).

We have estimated the helium mass accretion rate to be several 
times $10^{-7} M_\sun$~yr$^{-1}$. In such a high accretion rate, 
helium flash is very weak and only a part of the envelope blown in the wind
and the rest of them accumulates on the WD \citep{kat99}. Therefore, the 
WD will grow in mass after many cycles of helium shell flashes.
The fate of the WD is either a Type Ia supernova or an 
accretion induced collapse to a neutron star, depending on
its initial WD mass \citep{nom91}.

The nature of the companion star is also interesting.  
The companion could be a helium star or a helium WD, from which the 
WD accretes helium matter directly. 
Another possibility of the companion is 
a helium rich main-sequence star as suggested in the companion star of 
U Sco. With such high accretion rates, 
hydrogen shell flashes are very weak or the shell burning is almost 
stable. For example, hydrogen shell burning is steady for 
mass accretion rate higher than 
$3.8 \times 10^{-7} M_\sun$~yr$^{-1}$ for $X=0.7$ and 
$6.2 \times 10^{-7} M_\sun$~yr$^{-1}$ for $X=0.5$ for
$1.35~M_\sun$ WD. These value depends very weakly on the WD mass. 
This means that for the mass accretion rate  we 
have estimated for V445 Pup, hydrogen burning is stable for $X=0.7$ but
unstable shell flashes repeat for $X=0.5$. For $1.33~M_\sun$ WD, 
hydrogen burning is stable for $X\ge 0.5$ but unstable for $X=0.35$. 
In case of steady hydrogen burning, 
the brightness of the WD is estimated to be $M_V = 7.8$ for $1.35~M_\sun$ 
WD, and 7.6 for $1.33~M_\sun$ WD.  These values are 
below the upper limit of observation. 
Before the outburst, V445 Pup shows
a constant brightness of $\sim 14.5$ mag. This luminosity may 
be a contribution from the accretion disk, not included in our theoretical
light curves.

To summarize our results: 
 We have calculated light curves of helium novae 
and reproduce observational data for V445 Puppis.
From light curve modeling, we conclude that 
the WD mass is more massive than $1.3 M_{\odot}$, 
the envelope mass at ignition is several times $10^{-5} M_\sun$, 
the mass accretion rate from the companion is several times 
$10^{-7} M_\sun$~yr$^{-1}$, 
and the recurrence period of helium nova outbursts is several tens of years. 
Distance to the star is estimated to be 640-700 pc with $A_V=0.78$.   
Because the wind is weak, the WD is growing in mass, and therefore, V445 Pup 
is a candidate of Type Ia supernova progenitors.

\acknowledgments
  We are grateful to VSNET members who contributed to V445 Pup light curve.
This work was supported in part by the Grants-in-Aid from
The 21st Century COE (Center of Excellence) program
( Research Center for Integrated Science) of the Minister of Education,
Culture, Sports, Science, and Technology, Japan.

\begin{figure}
\plotone{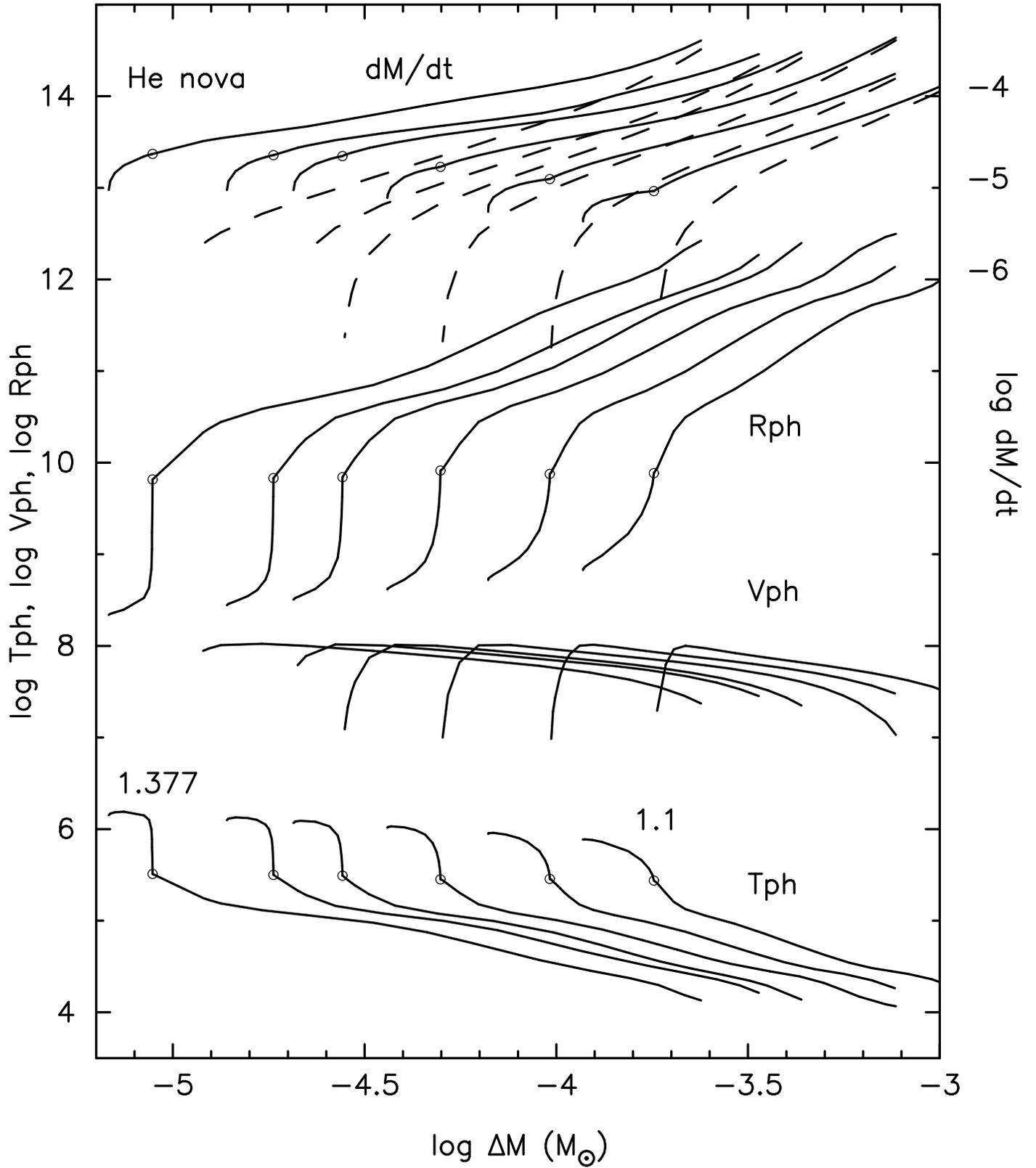}
\caption{ {\it Top}: Total envelope mass decreasing rate 
(nuclear burning + wind mass loss) (solid curve),
and wind mass loss rate (dashed curve) in units of $M_\sun$~yr$^{-1}$, 
{\it Second}: Photospheric radius $R_{\rm ph}$~(cm), 
{\it Third}: wind velocity $V_{\rm ph}$~(cm~s$^{-1})$, 
{\it Bottom}: temperature $T_{\rm ph}$~(K), against
the envelope mass ($\Delta M$) in units of $M_\sun$. 
Time runs from right to left because the envelope 
mass is decreasing in time. The wind mass loss 
stops at the point marked by a small
open circle. The WD mass is 
1.1, 1.2, 1.3, 1.33, 1.35, and 1.377 $M_\sun$, from right to left.
  \label{fig1}}
\end{figure}

\begin{figure}
\plotone{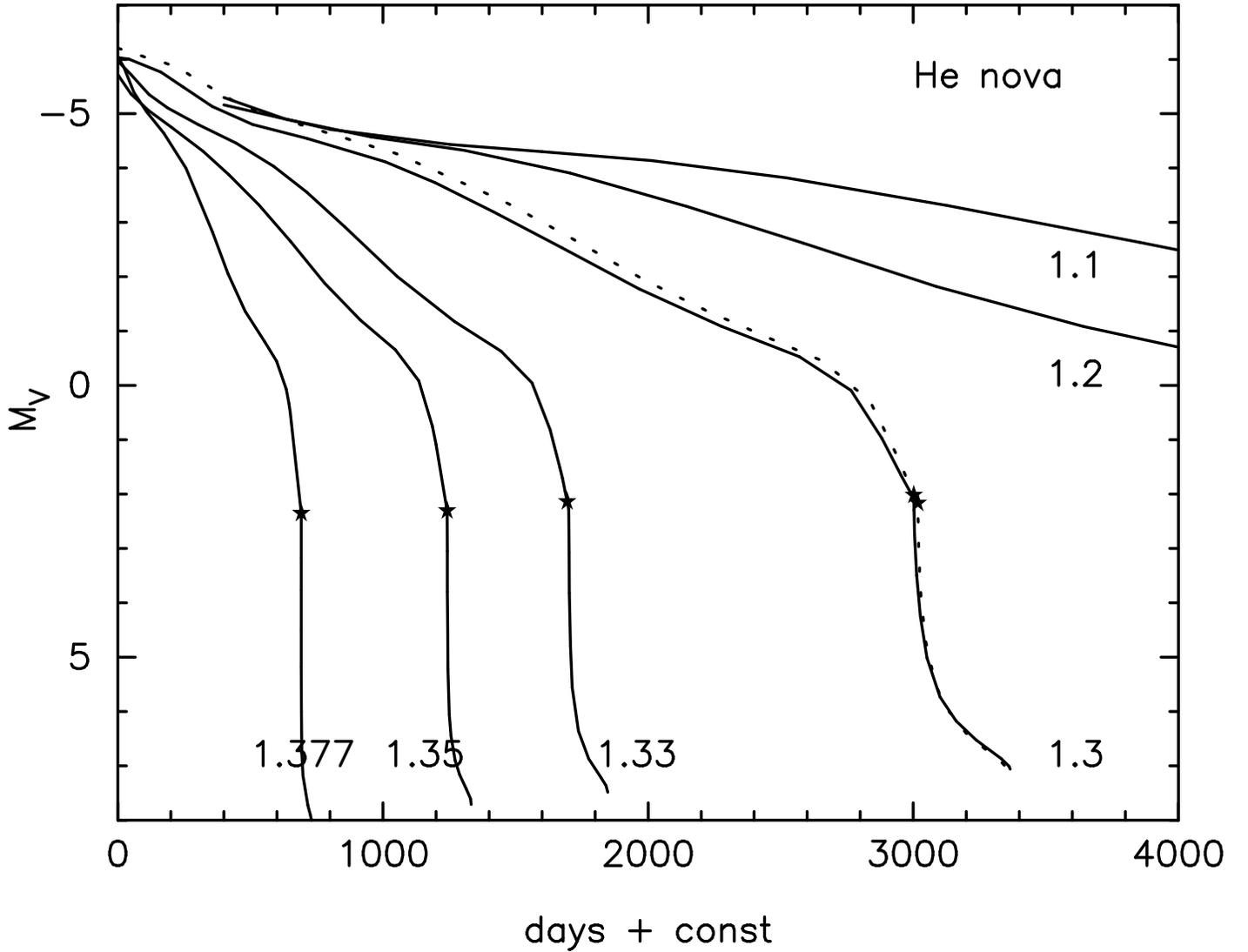}
\caption{ Template of our theoretical light curves for helium novae
with the composition of $(X, Y, C+O, Z)=(0.0, 0.48, 0.5, 0.02)$. 
WD masses are attached to each light curve in units of $M_\sun$. 
The dotted curve denotes a 1.3 $M_{\odot}$ WD model with different 
composition of $(X, Y, C+O, Z)=(0.0, 0.38, 0.6, 0.02)$.  The wind mass loss 
stops at asterisk on each curve.
  \label{fig2}}
\end{figure}

\begin{figure}
\plotone{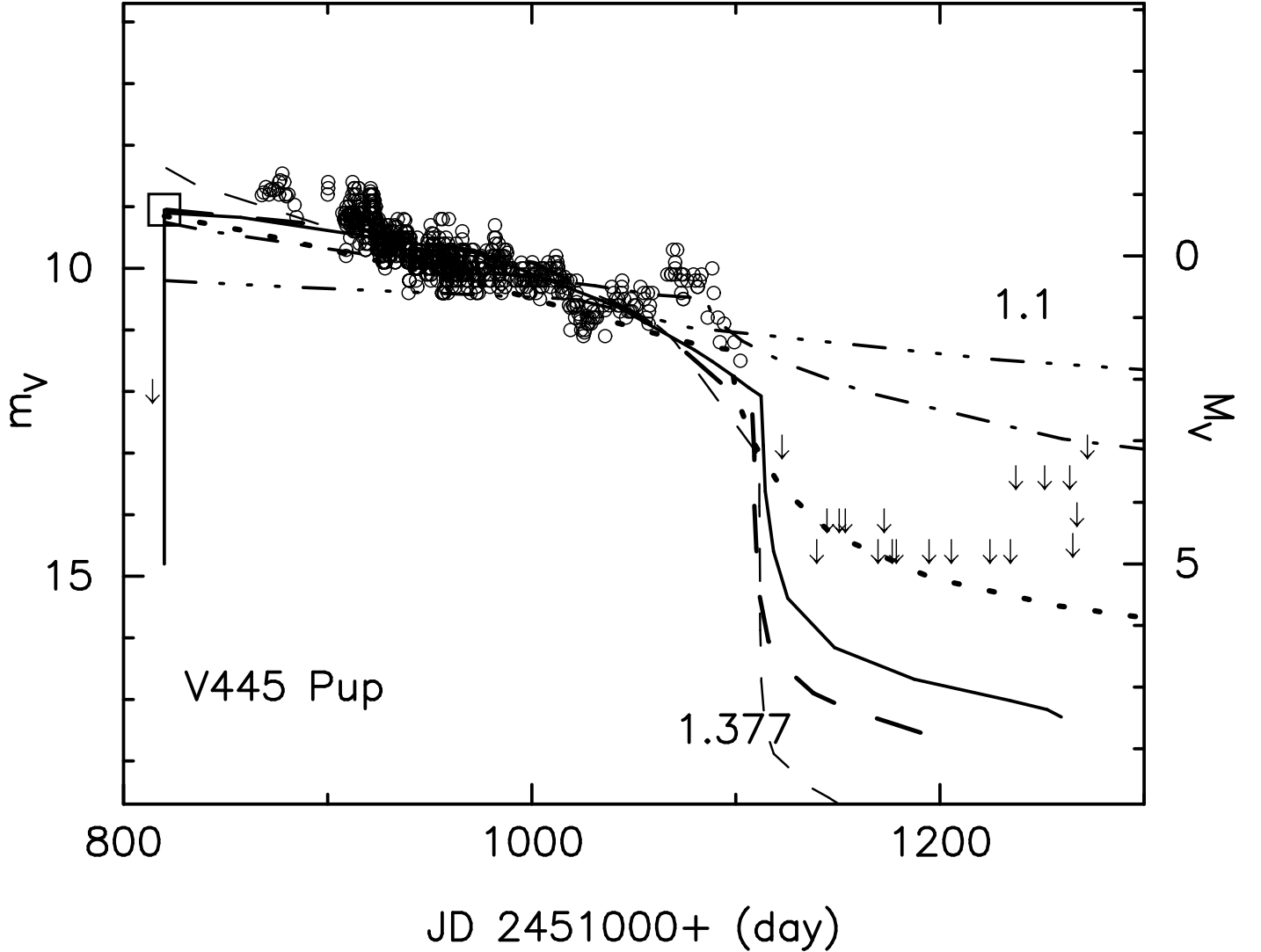}
\caption{ Light curve modeling of V445 Pup for
1.1 ({\it dash-three-dotted}) 
1.2 ({\it dash-dotted}) , 1.3 ({\it dotted}), 
1.33 ({\it solid}), 1.35 ({\it dashed}), and
$1.377~M_\sun$ ({\it thin dashed}) WDs. 
The apparent ($m_V$) and absolute ($M_V$) magnitude scale are shown 
in the left- and right-hand side, respectively.
Small open circles represent observational data while 
arrows indicate upper limits, both of which are 
taken from VSNET archive.
Only a later part of the 
theoretical curve in Fig. 2 is well fitted with the V445 Pup data. 
The theoretical curves except $1.33~M_\sun$ WD
are shifted in the vertical direction (magnitude
scale is in write-hand side): 1.5, 1.5 and 0.5 mag
upward for 1.1, 1.2 and $1.3~M_\sun$ WDs, and 
0.2 and 0.9 mag downward for 1.35 and
$1.377~M_\sun$ WDs, respectively.  
  \label{fig3}}
\end{figure}

 

\end{document}